 \documentstyle[12pt]{ws-art}

\begin{document}

\title{
HADRON-HADRON INTERACTIONS
 \footnote{
Invited talk  presented at the
Conference on Physics with GeV--Particle Beams,
J\"ulich, Germany, August 22-25, 1994.
}
}
\author{
K. Holinde \\
{\sl Institut f\"ur Kernphysik (Theorie), Forschungszentrum J\"ulich GmbH,} \\
{\sl D-52425 J\"ulich, Germany }}
\maketitle

\begin{abstract}

The present status of the chiral approach to the $NN$ interaction as
proposed by Weinberg is discussed. The important role of correlation
effects and explicit vector meson ($\rho$, $\omega$) exchange in the dynamics
of baryon-baryon interactions is demonstrated. As an example, the
inclusion of the exchange of a correlated pair of $\pi$ and $\rho$ meson
between two nucleons appears to be mandatory in order to resolve a
long-standing puzzle concerning the formfactor at the
pion-nucleon-nucleon vertex.

\end{abstract}

\section{
Introduction
}

Quantum chromodynamics (QCD) is the underlying theory of strong
interactions, with quarks and gluons as fundamental degrees of freedom.
However, in the non-perturbative region of low and medium energy
physics, mesons and baryons definitely keep their importance as
efficient, collective degrees of freedom for a wide range of nuclear
phenomena.
There is a widespread view that a successful effective field theory in
terms of hadronic degrees of freedom should have the same symmetries
as QCD. Chiral symmetry, in the limit of vanishing quark masses, is an
exact QCD symmetry; chiral perturbation theory\cite{1} is an effective
field theory in terms of the light pseudoscalar mesons ($\pi$,$K$,$\eta$),
the Goldstone bosons of spontaneously broken chiral symmetry. It is
important to realize that such an effective field theory is
mathematically eqivalent to QCD. At sufficiently low energies,
smaller than the QCD mass scale of about 1GeV, it leads to a systematic
expansion of effective Lagrangians and scattering amplitudes in powers
of momenta and quark masses (the latter arise from the symmetry breaking
terms).

Of course, the effectiveness (as a fastly converging scheme) of chiral
perturbation theory is by definition restricted to quite low energies.
Moreover,
a perturbational method can only work if the interactions are weak, which is
the
case for reactions involving the Goldstone bosons as external particles. Given
these constraints, chiral perturbation theory is remarkably successful; for a
review, see e.g.\ the recent article by Meissner\cite{2}.

On the other hand, for baryon-baryon and baryon-antibaryon interactions strong
correlation effects appear already at threshold; it is hard to get such effects
from an expansion which, for practical reasons, has to be limited to low
orders. Furthermore, higher-mass mesons, especially vector mesons $\rho$ and
$\omega$, play an important role.  In my opinion, a successful description
requires to treat both dynamical mechanisms explicitly, as will be discussed
below.

\section{
Chiral Dynamics in the nucleon-nucleon system.
}

Let me start by looking at present chiral treatments of the nucleon-nucleon
($NN$) system. Recently, quite a lot of papers have appeared
dealing with this issue \ctt{3}{6}. Whereas Refs.~\ctt{3}{5} concentrated
on the chiral structure of a specific contribution to the $NN$ interaction,
 namely $2\pi$-exchange, Ref.~\cite{6} presented a ``complete'' $NN$ potential
strictly obeying chiral symmetry. It is based on the most general
effective chiral Lagrangian involving low momentum pions, nonrelativistic
nucleons and $\Delta$-isobars, and is considered up to third order ($(Q/M)^3$,
with $Q$ a typical external momentum and $M$ the QCD mass scale) in the
chiral expansion, thus containing tree as well as one-loop diagrams.

\begin{figure}[ht]
\vskip 08.5cm
\includegraphics{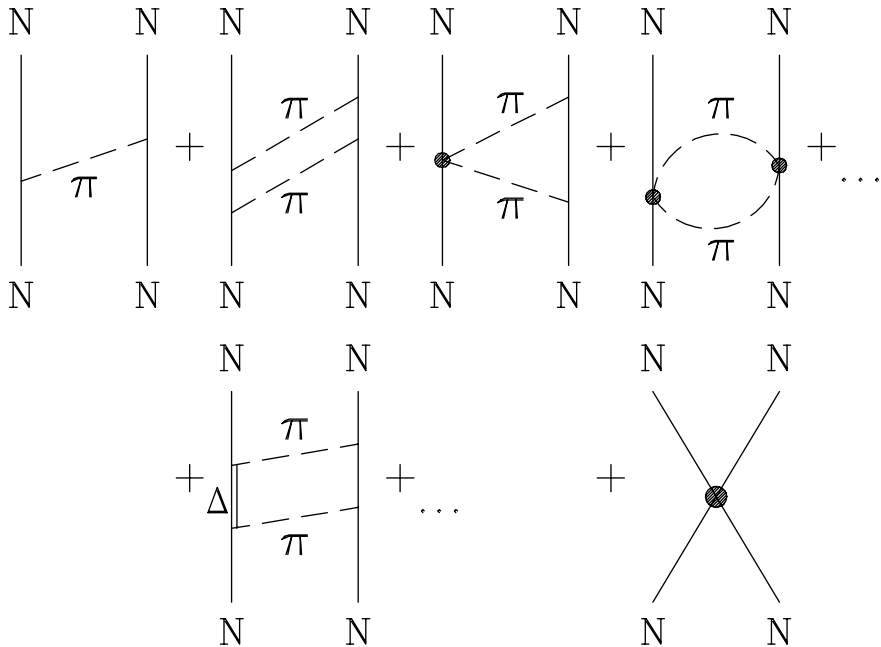}
\caption{\label{figI}
Diagrams included in the chiral approach of Ref.~\protect\cite{6}.
}
\end{figure}

\begin{figure}
\vskip 16.75cm
 \includegraphics{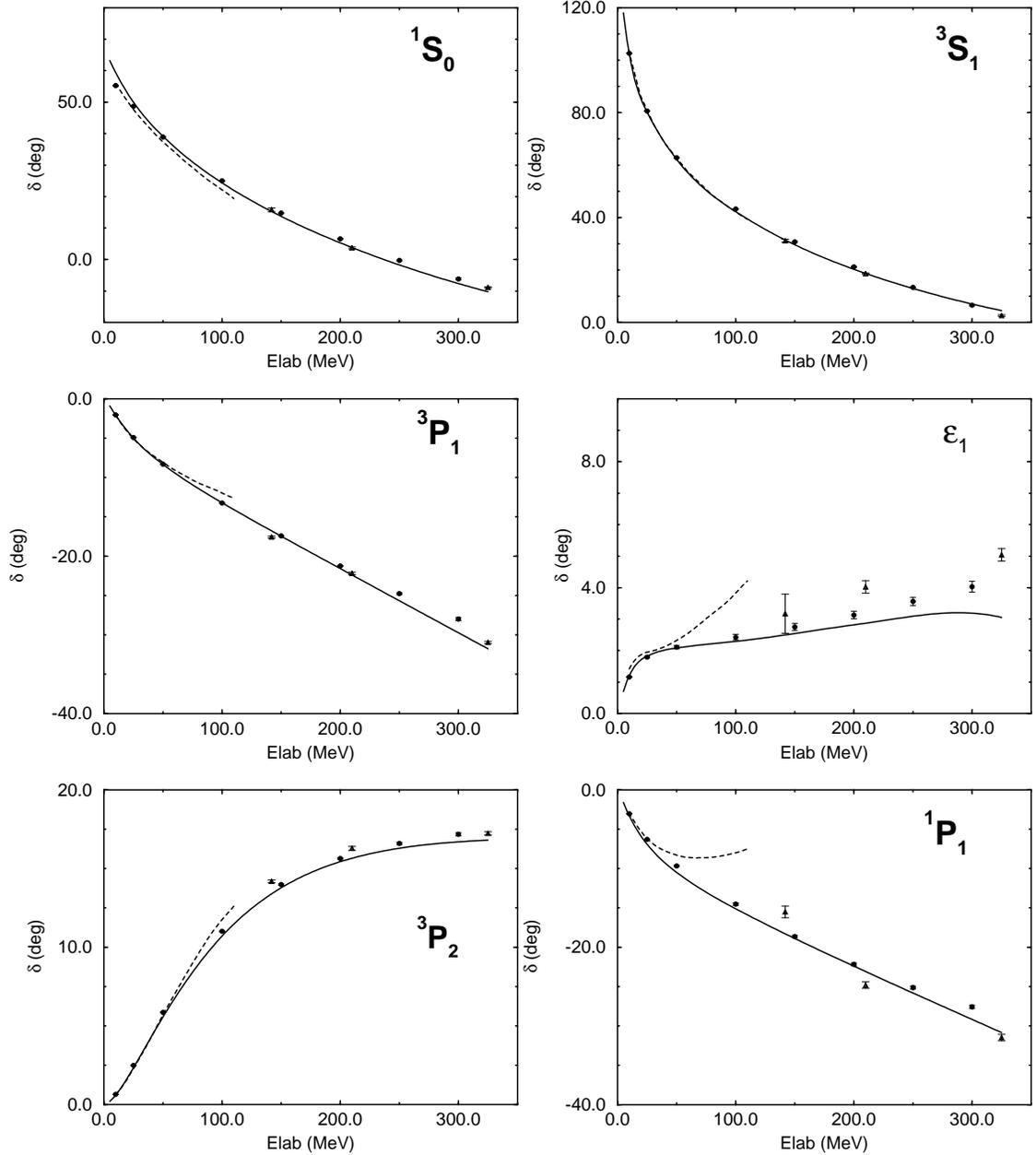}
\caption{\label{figII} $NN$ phase shifts, in selected partial waves, as a
function of the nucleon lab. energy.  The solid lines give the predictions
derived from the Bonn potential (Ref.\ \protect\cite{7}) whereas the dashed
lines, taken from Ref.~\protect\cite{6}, originate from the chiral approach.
The
experimental error bars are taken from Refs.~\protect\cite{a,b} }
\end{figure}

In diagrammatic terms, the resulting $NN$ potential has the structure shown in
Fig.~\ref{figI}. It consists of one-pion-exchange (OPE) and $2\pi$-exchange
terms with $N$ and $\Delta$ intermediate states.  There are additional
$2\pi$-exchange processes in the chiral approach (Ref.~\cite{6}) involving the
$N\bar N\pi\pi$ vertex (see Fig.~\ref{figI}), whose presence is dictated by
chiral symmetry. However, due to partial cancellations (Refs.~\cite{5,4}) they
are actually quite small, much smaller e.g.\ than ``$\sigma$''- exchange
representing correlated $2\pi$-exchange in the scalar-isoscalar channel, which
provides the main part (about $2/3$, see Ref.~\cite{7})
 of the $NN$ intermediate range attraction.  The chiral approach absorbs this
contribution together with other important pieces involving vector
($\rho,\omega$) mesons into the contact terms, i.e. terms of zero range (last
diagram of Fig.~\ref{figI}).  In order to solve the corresponding Schroedinger
equation a Gaussian cutoff function, $exp(-Q^2/L^2)$, with $L = 3.9 fm^{-1}$,
has been used. Results for some low ($S$ and $P$) $NN$ partial wave phase
shifts
are shown in Fig.~\ref{figII} (taken from Ref.~\cite{6}), in comparison to
those
obtained from the (full) Bonn potential\cite{7}.  In addition, table 1 shows
some deuteron parameters.

\begin{table}\small
\caption{\label{tableI} Deuteron properties predicted by the
chiral model of Ref.~\protect\cite{6}, the Bonn $NN$ potential
(Ref.~\protect\cite{7}), and the Bonn potential with a soft $\pi NN$ form
factor
($\Lambda_{\pi NN}= 1GeV$) plus correlated $\pi\rho$ exchange
(Ref.~\protect\cite{21}) as discussed in Section~5.  For the experimental
values, see Refs.~\protect\cite{7,22}.  }
\begin{center}
\begin{tabular}
{l|cccc}
              & experiment  & chiral model     &    Bonn      &   modified Bonn
\cr
              &             & (Ref.~\protect\cite{6}) &(Ref.~\protect\cite{7})
&
\cr
\noalign{\hrule}
Binding energy (MeV)        &  2.2245754        & 2.18   & 2.2247 & 2.2246 \cr
Quadrupole  moment ($fm^2$) & 0.2859$\pm0.0003$ & 0.231  & 0.2807 & 0.2791 \cr
asymptotic D/S ratio        & 0.0256$\pm0.0004$ & 0.0239 & 0.0267 & 0.0266 \cr
                            &(0.0271$\pm0.0008$)&        &        &
\end{tabular}
\end{center}
\end{table}

In this first confrontation of a chiral $NN$ potential \`a la Weinberg to the
very precise $NN$ data rough agreement is reached in the low-energy domain.
Note
that this model has 26 parameters available (mainly the coefficients in the
contact terms) compared to about 10 meson-baryon coupling constants and
formfactor parameters in the Bonn potential. A combined look at the
$\varepsilon_1$-parameter and deuteron parameters ($Q$, $D/S$) is instructive
since both measure the size of the $NN$ tensor force.  Obviously, in the
present
chiral model, the energy dependence of this piece is not realistic since the
deuteron values are too low whereas the $\varepsilon_1$-values grow too fast
with increasing energy. There are two possible sources for this discrepancy:
First, the chiral expansion introduces a polynomial momentum dependence, with
adjustable parameters, at the $\pi NN$ vertex; it should be checked by the
authors of Ref.~\cite{6} whether their parametrization chosen is in reasonable
agreement with the known monopole structure of the $\pi NN$ formfactor, see
Section 4.  Furthermore, $\rho$-exchange provides an important contribution to
the tensor force (see Section 3) reaching out well into intermediate $NN$
distances of $1 - 2 fm$. It is highly questionable whether this piece (as well
as ``$\sigma$'' and $\omega$ exchange) can be sufficiently represented by
zero-range contact terms.

In my opinion, in present chiral approaches to the $NN$ system, the price you
pay (having to treat important contributions in a very rough way) is too high
for what you get (a strict chirally symmetric $V_{NN}$ derived from a
systematic
expansion).

\section{
Structure of $\pi$ and $\rho$ exchange in the $NN$ system
}

It is well known that the physics of ``$\sigma$'' and $\rho$ exchange, so
important in the $NN$ system, is mainly built up by correlated $2\pi$-exchange.
(For a recent dynamical model suitable for the discussion of medium
modifications, see the paper by Kim et al.\cite{10}.)  Therefore, in principle,
the chiral approach can take such effects, at least to some extent, into
account
by going to still higher orders in the chiral expansion, at the expense of
getting additional open parameters.

\begin{figure}[ht]
\vskip 7cm
\includegraphics{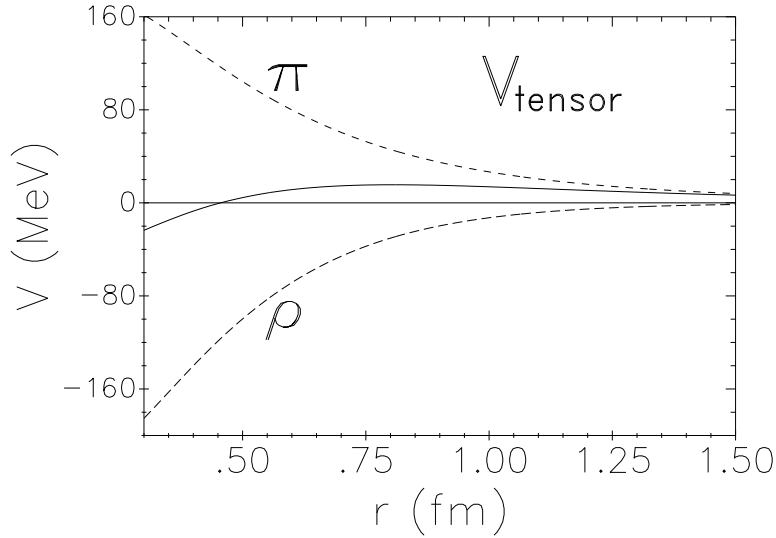}
\caption{
\label{figa}
$NN$ tensor force $V_T(r)$ as function of the two-nucleon distance
$r$, due to $\pi$ and $\rho$ exchange. The solid line denotes the sum of both
contributions. The parameter values are taken from OBEPR
(Ref.~\protect\cite{7}).
}
\end{figure}

The importance of $\rho$-exchange for the dynamics of the $NN$ system derives
from the following fact: It provides a sizable tensor force, which has opposite
sign to the tensor force generated by one-pion-exchange, see Fig.~\ref{figa}.
Thus there is a strong cancellation, over a relatively broad range of energies
and distances, between $\pi$ and $\rho$ exchange in the tensor channel. A
similar cancelation occurs in the strange sector, between $K$ and $K^*$
exchange, e.g.\ in the hyperon-nucleon interaction.  
Therefore, in the $NN$ (and in the baryon-baryon system in general) it is
strongly suggested to always group $\pi$ and $\rho$ (as well as $K$ and $K^*$)
together in order to reach sufficient convergence in the expansion of the
irreducible kernel (potential).
 To single out $\pi$-exchange as done in the chiral approach might be more
systematic from a formal point of view; on the other hand, by putting
$\rho$-exchange into the contact terms, one essentially loses the strong
convergence-generating mechanism at intermediate distances provided by
$\rho$-exchange.  Thus, from a physical standpoint, it appears mandatory to
treat $\pi$ and $\rho$ exchange on an equal footing.

\begin{figure}
\vskip 4.5cm
\includegraphics{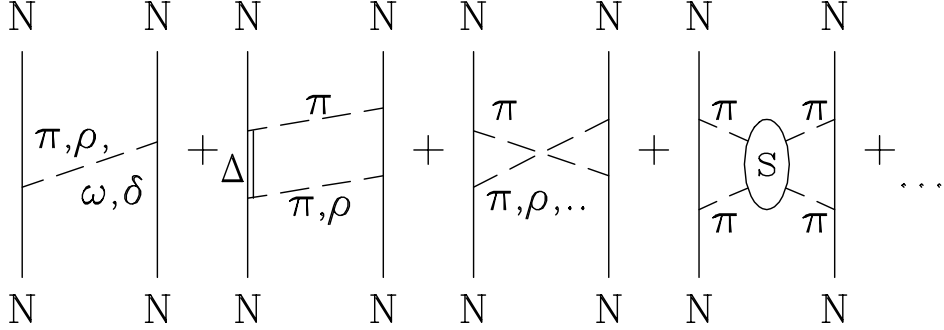}
\caption{\label{figVIII}
Diagrams included in the Bonn $NN$ potential\protect\cite{7}.}
\end{figure}

In fact, this procedure has been an essential guideline when constructing the
Bonn potential. Unfortunately (though for understandable reasons, see below) it
was not followed to a sufficient degree: Whereas, in second-order diagrams
(cp.\
Fig.~\ref{figVIII}) $\pi\pi$ as well as $\pi\rho$ exchange have been included
for uncorrelated processes (with $N$ and $\Delta$ intermediate states) this has
not been done for correlated processes: correlated $2\pi$-exchange processes
have been included (in terms of sharp-mass $\sigma'$ and $\rho$ exchange) but
correlated $\pi\rho$ processes have been left out. The reason is quite simple:
The evaluation of this missing piece is technically quite complicated, much
more
involved (due to the spin of the $\rho$) compared to correlated
$2\pi$-exchange. More importantly, a dynamical model for the interaction
between
a $\pi$ and a $\rho$ meson was not available.

At the time of the construction of the Bonn potential (about 10 years
ago) the omission of this piece did not seem to be serious. $NN$ scattering
data in a broad energy range as well as the deuteron data are described
quantitatively with the full Bonn model. So why care about a missing
piece whose practical relevance was not at all obvious? Soon after the
publication of the Bonn potential in 1987 it became however clear that
omission of correlated $\pi\rho$ exchange demanded a high price, and this
has to do with the structure of the $\pi NN$ vertex.

\section{ The $\pi NN$ vertex }
\begin{figure}[ht]
\vskip 4.50cm
\includegraphics{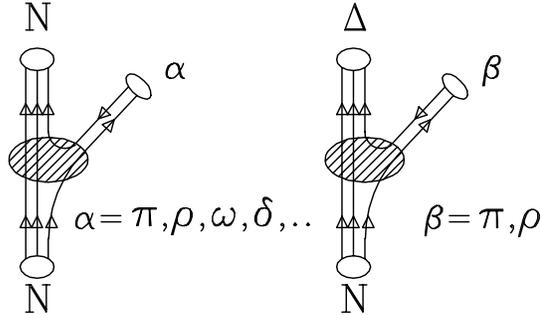}

\caption{\label{figIX} Baryon-baryon-meson couplings included in the Bonn $NN$
potential (Ref.~\protect\cite{7}).  }
\end{figure}

Basic ingredients of the Bonn meson exchange $NN$ model\cite{7} are the
meson-baryon couplings or vertex functions visualized in Fig.~\ref{figIX}.
These
vertex functions are the ``elementary'' building blocks of an effective and
consistent meson exchange description based on QCD.

The analytic structure of these couplings is (essentially) determined by the
quantum numbers of the particles involved at the vertex. The strength is
parametrized by coupling constants $g_{B'B\alpha}$; in addition, formfactors
with cutoff masses $\Lambda_{B'B\alpha}$ as additional parameters are included,
which parametrize the effect of the hadron extension and therefore represent a
truly physical concept. In general, the formfactor depends on all four-momenta
of the particles involved at the vertex, i.e.\ $F_{B'B\alpha} =
F(q_{B'}^2,q_B^2,q_\alpha^2)$ \ and is normalized to 1 when all particles are
on
the mass shell.

While ultimately these formfactors have to be derived from QCD they are
for the moment parametrized, usually in monopole form, keeping the
dependence on the momentum of the exchanged particle only. With the
additional suppression of the dependence on $q_\alpha^0$, the formfactors
relevant in the $NN$ system can then be written as

\begin{equation}
F_{B'B\alpha} = {\Lambda_{B'B\alpha}^2 - m_\alpha^2 \over \Lambda_{B'B\alpha}^2
+ \vec q_\alpha^2} \ .
\end{equation}

The cutoff masses $\Lambda_{B'B\alpha}$ together with the coupling constants
$g_{B'B\alpha}$ represent the only parameters in the Bonn potential; they have
been adjusted to the $NN$ data. For the $\pi NN$ cutoff mass,
$\Lambda_{NN\pi}$,
the resulting value turned out to be rather large ($=1.3 GeV$) leading to a
mild
suppression of OPEP in the inner region ($r\le 1fm$) only, and thus to a hard
formfactor. This is necessary in order to have sufficiently strong tensor force
to reproduce the deuteron properties, especially the asymptotic $D$- to
$S$-wave
ratio and the quadrupole moment\cite{14}.

However, there is a long-standing discrepancy between this rather large value
required in present-day potential models and information from other sources;
the
latter consistently point to a much smaller value for $\Lambda_{NN\pi}$ around
$0.8 GeV$\cite{15}. In fact, a recent lattice calculation\cite{16} of the $\pi
NN$ formfactor confirms this result, finding a monopole mass of $(0.75\pm0.14)
GeV$. Such a soft formfactor leads to a strong suppression of the OPE tensor
force already at intermediate distances, in apparent disagreement with the
deuteron data. It is sometimes argued that a reduction (or even elimination) of
$\rho$ exchange in $NN$ models would remedy the situation easily since the
tensor force of $\rho$ exchange has opposite sign compared to
$\pi$-exchange. This can certainly be done if one looks at the deuteron channel
only. However the sizable strength of $\rho$ exchange is tightly constrained by
informations from $\pi N$ scattering via dispersion theory (see Ref.~\cite{10})
and in fact required by the precisely measured triplet $P$-wave $NN$ phase
shifts; it therefore cannot be arbitrarily changed.

\begin{figure}[ht]
\vskip 4.5cm
\includegraphics{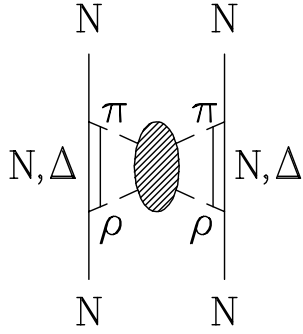}
\caption{\label{figX}
Model for the correlated $\pi\rho$ exchange as evaluated in
Ref.~\protect\cite{21}.}
\end{figure}

Consequently we have to find mechanisms so far not included in meson exchange
$NN$ models which provide additional (short ranged) tensor force.  Two years
ago
it was pointed out by Ueda\cite{17} that the contribution arising from the
exchange of a correlated three-pion state between the nucleons, with the
quantum
numbers of a pion, should enhance the short-range tensor part of the $NN$
interaction. Correlated $\pi\rho$ exchange addressed in the foregoing chapter
is
a good part of it, and there is obviously no way around this contribution.

\section{ Correlated $\pi\rho$ exchange in the $NN$ system }
\label{pirho}
The explicit evaluation of the diagram in figure~\ref{figX} requires as input a
realistic $\pi\rho$ $T$-matrix. In the absence of elastic $\pi\rho$ scattering
data (due to the fast decay of the $\rho$ meson into 2 pions) we have to rely
completely on a dynamical model.

\begin{figure}[ht]
\vskip 3.75cm
\includegraphics{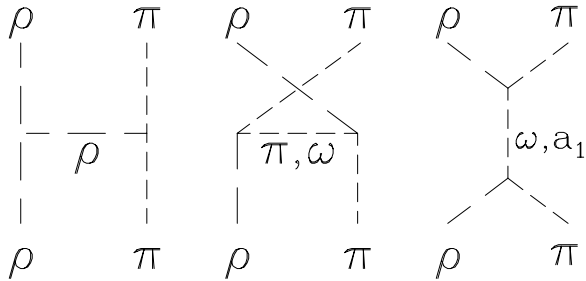}
\caption{\label{figXI}
Driving terms for the $\pi\rho$ interaction model of
Refs.~\protect\cite{21,18}.}
\end{figure}

\begin{figure}[ht]
\vskip 7cm
\includegraphics{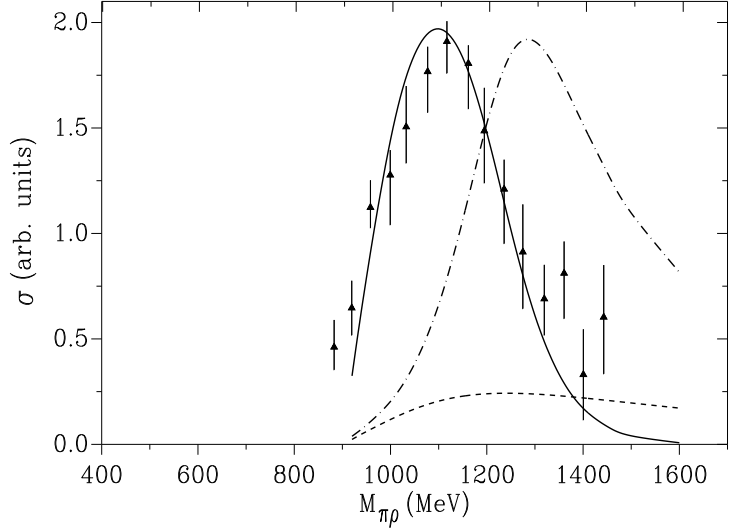}
\caption{\label{figXII} Mass spectrum in the $a_1$ channel obtained from $\pi
p\to 3\pi n$ (Ref.~\protect\cite{20}).  The solid line shows the prediction of
our full model whereas the dashed line denotes the result of the non-pole part
of the $T$-matrix, $T_{np}$, only.  The dash-dotted line provides the result of
the pole part $T_p$ only, with readjustment of the parameters to keep the same
pole values}
\end{figure}

We\cite{18} have recently constructed a corresponding potential model, in
complete analogy to the $\pi\pi$ case (Ref.~\cite{11}), with driving terms
shown
in Fig.~\ref{figXI}. Open parameters are adjusted mainly to empirical
information in the $a_1$-channel (the $a_1$ decaying into $\pi$ and $\rho$),
namely the resonance parameters (pole position) of the $a_1$-particle\cite{19}
($m_{a_1} = 1.26 GeV$, $\Gamma_{a_1} = 0.4 GeV$) obtained from $\tau$-meson
decay, and the mass spectrum obtained from charge exchange production\cite{20},
$\pi p\to 3\pi n$, see Fig.~\ref{figXII}. Note that the shift of the maximum of
the mass spectrum away from the true pole position is due to the sizable
non-pole amplitude $T_{np}$, obtained by iterating $V_{np}$ (the first two
diagrams of Fig.~\ref{figXI}.).

This non-pole contribution now acts in all other channels, e.g.\ in the
pionic channel of interest here, with a definite strength distribution,
which is characteristic of the underlying model.

This $\pi\rho$ $T$-matrix is now inserted into the correlated $\pi\rho$
exchange
diagram of Fig.~\ref{figX}. The evaluation\cite{21} proceeds via the same
dispersion-theoretic treatment as used for the $\pi\pi$ case in
Ref.~\cite{10}. The result can be represented as integral over various spectral
functions, the number of which depend on the considered channel. For the pionic
channel we have one spectral function $\rho^{(\pi)}$: \begin{equation} %
V_{\pi\rho}^{(\pi)} \ \sim \ \int\limits_{(m_\pi+m_\rho)^2}^\infty \
{\rho^{(\pi)}(t')\over
                                                                 t'-t }\    dt'
{}.
\end{equation}

\begin{figure}
\vskip 7cm
\includegraphics{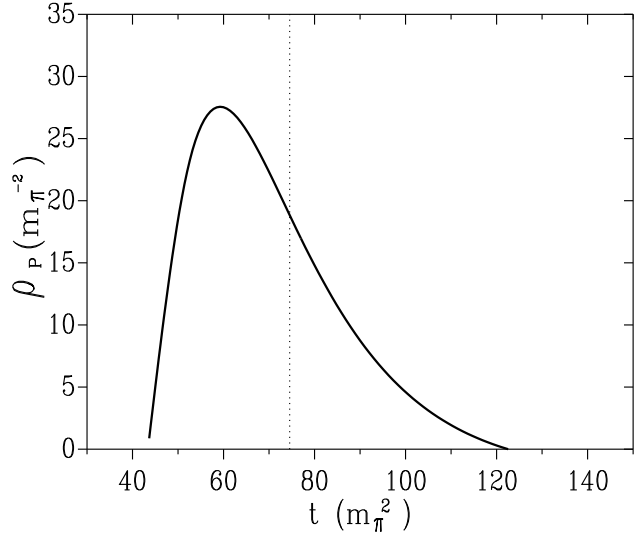}

\caption{ \label{figXIII} Spectral function $\rho^{(\pi)}$ as a function of
$t'$
in the pseudophysical region $t'\ge (m_\pi+m_\rho)^2$, characterizing the
correlated $\pi\rho$ exchange contribution to the $NN$ interaction in the
pionic
channel. The vertical line indicates the $\delta$ function at
$m_{\pi'}=1.2GeV$,
representing sharp mass $\pi'$ exchange as used in Ref.~\protect\cite{22}.}
\end{figure}

$\rho^{(\pi)}$ is shown in Fig.~\ref{figXIII}. Obviously the correlated part
provides a sizable contribution, with a peak around $1.1 GeV$, somewhat smaller
than the mass ($1.2 GeV$) of the phenomenological $\pi'$ introduced in some
recent $NN$ models to accommodate a soft $\pi NN$ formfactor\cite{22}.

\begin{figure}
\vskip 7cm
\includegraphics{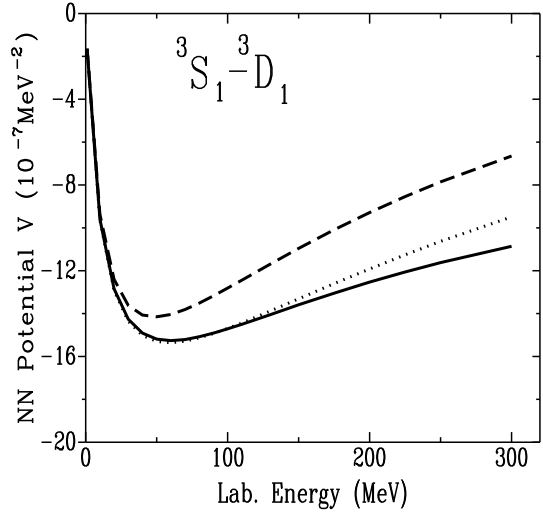}
\caption{ \label{figXIV} On-Shell $NN$ potential $V(q,q)$ as function of the
nucleon lab. energy, $E_{lab}={q^2/2M_N}$ in the ${^3D_1}\to{^3S_1}$
transition. The dotted line denotes the one-pion exchange potential as used in
the Bonn potential, with $\Lambda_{\pi NN}=1.3 GeV$.  For the dashed line,
$\Lambda_{\pi NN}=1 GeV$.  The solid line results if correlated $\pi\rho$
exchange (Fig.~\protect\ref{figX}), in the pionic channel, is added to the
dashed line.}
\end{figure}

Indeed, as shown in Fig.~\ref{figXIV} and Table 1, the resulting interaction
due
to correlated $\pi\rho$ exchange in the pionic channel is able to
counterbalance
the substantial suppression induced in OPEP when going from a cutoff mass
$\Lambda_{NN\pi}$ of $1.3 GeV$, phenomenologically required in the (full) Bonn
potential, to a value of $1.0 GeV$. Further tensor contributions, with the same
sign as pion exchange, arise from correlated $\pi\rho$ exchange in the
$a_1$-channel. If these will be included, too, $\Lambda_{NN\pi}$ can be
decreased further, into the region of $0.8 GeV$. Thus the inclusion of
correlated $\pi\rho$ exchange in a meson exchange $NN$ model should allow to
use
the correct, soft $\pi NN$ formfactor, and thus resolve a long-standing puzzle.

\section{
Concluding remarks
}

In this talk, I have tried to convince you that, in low energy baryon-baryon
interactions, chiral symmetry does not appear to be the dominant symmetry: A
meson exchange model like the Bonn potential, which violates (to some extent)
chiral symmetry, is able to describe the data quantitatively.  On the other
hand, a model strictly obeying chiral symmetry but restricting itself to pions
as mesonic degrees of freedom and one-loop contributions shows noticeable
deficiencies in the description of the empirical situation, already at very low
energies. A minimal requirement for improvement would be the inclusion of
two-loop contributions, i.e.\ to go to still higher order in the chiral
expansion, in order to take at least part of the $2\pi$--correlation effects
(``$\sigma$'' and $\rho$ exchange) into account

Anyhow, there are a lot of baryonic processes whose physics is dominated by
quite high orders. For example, we have recently shown\cite{25} that the $\bar
pp\to \bar\Sigma\Sigma$ reaction goes predominantly through the
$\bar\Lambda\Lambda$ intermediate state, with strong initial ($\bar pp$),
intermediate ($\bar\Lambda\Lambda$), and final ($\bar\Sigma\Sigma$) state
interactions. Each interaction itself has sizable higher order contributions,
e.g.\ annihilation processes into 2 (or more) mesons. A treatment of these
coupled channels effects in the chiral approach, including systematically all
diagrams at the required order, is surely beyond our capabilities.

In this area of medium energy physics to be studied with our COSY accelerator,
it appears more efficient not to single out the Goldstone bosons as relevant
degrees of freedom but to include from the beginning higher mass mesons as
mediators of baryonic interactions, in this way summing up an important class
of
higher order correlation effects, as done already for the $\Delta$-isobar in
Ref.~\cite{6}. Further selected summations of processes to arbitrarily high
order, in a potential-type coupled channel framework, have to be
performed. Necessarily the full and exact implications of chiral symmetry are
then lost.

Clearly, as always in physics, the actual treatment of a given theory (kind of
approximations) depends on the phenomena under study and on the questions
addressed. If, for example, the main issue is to study modifications of
hadronic
interactions in the medium, chiral symmetry gains much more importance due to
chiral restoration in matter.  Thus, in order to study such effects seriously,
a
strictly chiral model is probably required. Still, one should always be aware
about what has been (necessarily?) sacrificed and about the consequences.

\vfil

\end{document}